\documentclass[twoside,twocolumn,9pt]{article}
\usepackage{extsizes}
\usepackage[super,sort&compress,comma]{natbib} 
\usepackage[version=3]{mhchem}
\usepackage[left=1.5cm, right=1.5cm, top=1.785cm, bottom=2.0cm]{geometry}
\usepackage{balance}
\usepackage{times,mathptmx}
\usepackage{sectsty}
\usepackage{graphicx} 
\usepackage{lastpage}
\usepackage[format=plain,justification=justified,singlelinecheck=false,font={stretch=1.125,small,sf},labelfont=bf,labelsep=space]{caption}
\usepackage{float}
\usepackage{fancyhdr}
\usepackage{fnpos}
\usepackage[english]{babel}
\addto{\captionsenglish}{%
  
}
\usepackage{array}
\usepackage{droidsans}
\usepackage{charter}
\usepackage[T1]{fontenc}
\usepackage[usenames,dvipsnames]{xcolor}
\usepackage{setspace}
\usepackage[compact]{titlesec}
\usepackage{hyperref}

\usepackage{subcaption}

\definecolor{cream}{RGB}{222,217,201}

\begin{document}

\pagestyle{fancy}
\thispagestyle{plain}
\fancypagestyle{plain}{

\fancyhead[C]{\includegraphics[width=18.5cm]{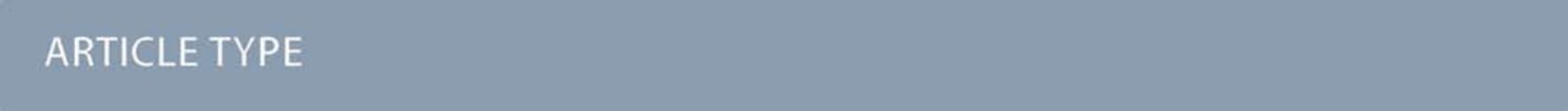}}
\fancyhead[L]{\hspace{0cm}\vspace{1.5cm}\includegraphics[height=30pt]{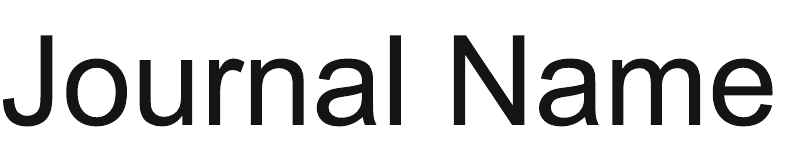}}
\fancyhead[R]{\hspace{0cm}\vspace{1.7cm}\includegraphics[height=55pt]{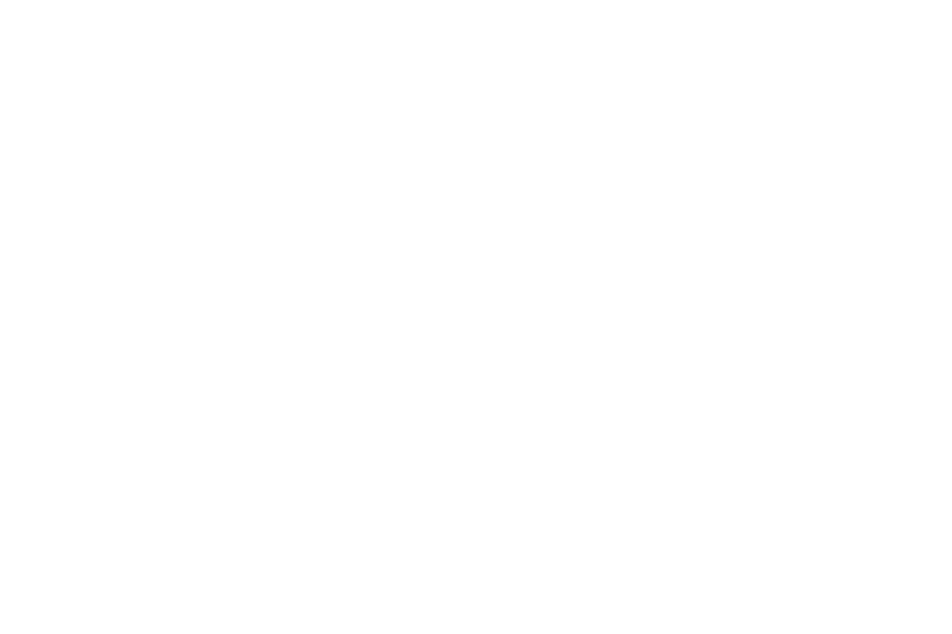}}
\renewcommand{\headrulewidth}{0pt}
}

\makeFNbottom
\makeatletter
\renewcommand\LARGE{\@setfontsize\LARGE{15pt}{17}}
\renewcommand\Large{\@setfontsize\Large{12pt}{14}}
\renewcommand\large{\@setfontsize\large{10pt}{12}}
\renewcommand\footnotesize{\@setfontsize\footnotesize{7pt}{10}}
\makeatother

\renewcommand{\thefootnote}{\fnsymbol{footnote}}
\renewcommand\footnoterule{\vspace*{1pt}%
\color{cream}\hrule width 3.5in height 0.4pt \color{black}\vspace*{5pt}} 
\setcounter{secnumdepth}{5}

\makeatletter 
\renewcommand\@biblabel[1]{#1}            
\renewcommand\@makefntext[1]%
{\noindent\makebox[0pt][r]{\@thefnmark\,}#1}
\makeatother 
\renewcommand{\figurename}{\small{Fig.}~}
\sectionfont{\sffamily\Large}
\subsectionfont{\normalsize}
\subsubsectionfont{\bf}
\setstretch{1.125} 
\setlength{\skip\footins}{0.8cm}
\setlength{\footnotesep}{0.25cm}
\setlength{\jot}{10pt}
\titlespacing*{\section}{0pt}{4pt}{4pt}
\titlespacing*{\subsection}{0pt}{15pt}{1pt}

\fancyfoot{}
\fancyfoot[LO,RE]{\vspace{-7.1pt}\includegraphics[height=9pt]{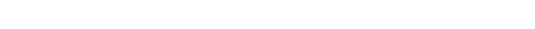}}
\fancyfoot[CO]{\vspace{-7.1pt}\hspace{13.2cm}\includegraphics{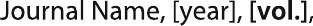}}
\fancyfoot[CE]{\vspace{-7.2pt}\hspace{-14.2cm}\includegraphics{head_foot/RF}}
\fancyfoot[RO]{\footnotesize{\sffamily{1--\pageref{LastPage} ~\textbar  \hspace{2pt}\thepage}}}
\fancyfoot[LE]{\footnotesize{\sffamily{\thepage~\textbar\hspace{3.45cm} 1--\pageref{LastPage}}}}
\fancyhead{}
\renewcommand{\headrulewidth}{0pt} 
\renewcommand{\footrulewidth}{0pt}
\setlength{\arrayrulewidth}{1pt}
\setlength{\columnsep}{6.5mm}
\setlength\bibsep{1pt}

\makeatletter 
\newlength{\figrulesep} 
\setlength{\figrulesep}{0.5\textfloatsep} 

\newcommand{\topfigrule}{\vspace*{-1pt}%
\noindent{\color{cream}\rule[-\figrulesep]{\columnwidth}{1.5pt}} }

\newcommand{\botfigrule}{\vspace*{-2pt}%
\noindent{\color{cream}\rule[\figrulesep]{\columnwidth}{1.5pt}} }

\newcommand{\dblfigrule}{\vspace*{-1pt}%
\noindent{\color{cream}\rule[-\figrulesep]{\textwidth}{1.5pt}} }

\makeatother

\twocolumn[
  \begin{@twocolumnfalse}
\vspace{3cm}
\sffamily
\begin{tabular}{m{4.5cm} p{13.5cm} }

\includegraphics{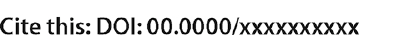} & \noindent\LARGE{\textbf{Numerical Simulations of Vorticity Banding of Emulsions in Shear Flows}} \\
\vspace{0.3cm} & \vspace{0.3cm} \\

 & \noindent\large{Francesco De Vita\textit{$^{a,*}$}, Marco Edoardo Rosti\textit{$^{b,a}$}, Sergio Caserta\textit{$^{c,**}$}, and Luca Brandt\textit{$^{a}$}} \\

\includegraphics{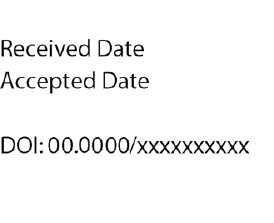} & \noindent\normalsize{Multiphase shear flows often show banded structures that affect the global behavior of complex fluids \emph{e.g.} in microdevices. Here we investigate numerically the banding of emulsions, \emph{i.e.} the formation of regions of high and low volume fraction, alternated in the vorticity direction and aligned with the flow (shear bands). These bands are associated with a decrease of the effective viscosity of the system. To understand the mechanism of banding experimentally observed we have performed interface resolved simulations of the two-fluid system. The experiments were performed starting with a random distribution of droplets which, under the applied shear, evolves in time resulting in a phase separation. To numerically reproduce this process, the banded structures are initialized in a narrow channel confined by two walls moving in opposite direction. We find that the initial banded distribution is stable when droplets are free to merge and unstable when coalescence is prevented. In this case, additionally, the effective viscosity of the system increases, resembling the rheological behavior of suspensions of deformable particles. Droplets coalescence, on the other hand, allows emulsions to reduce the total surface of the system and hence the energy dissipation associated to the deformation, which in turn reduces the effective viscosity.} \\

\end{tabular}

 \end{@twocolumnfalse} \vspace{0.6cm}

  ]

\renewcommand*\rmdefault{bch}\normalfont\upshape
\rmfamily
\section*{}
\vspace{-1cm}


\footnotetext{\textit{$^{a}$~Line\'e Flow Center and SeRC (Swedish e-Science Research Center), KTH Mechanics, S-100 44 Stockholm, Sweden.}}
\footnotetext{\textit{$^{b}$~Complex Fluids and Flows Unit, Okinawa Institute of Science and Technology Graduate University, 1919-1 Tancha, Onna-son, Okinawa 904-0495, Japan. }}
\footnotetext{\textit{$^{c}$~University of Napels "Federico II", Department of Chemical, Materials and Industrial Production Engineering, p.le Tecchio 80, 80125 Napoli, Italy. }}
\footnotetext{\textit{$^{*}$~E-mail: fdv@mech.kth.se}}
\footnotetext{\textit{$^{**}$~E-mail: sergio.caserta@unina.it}}




\section{Introduction}

\begin{figure}[ht]
  \centering
  \includegraphics[width=\linewidth]{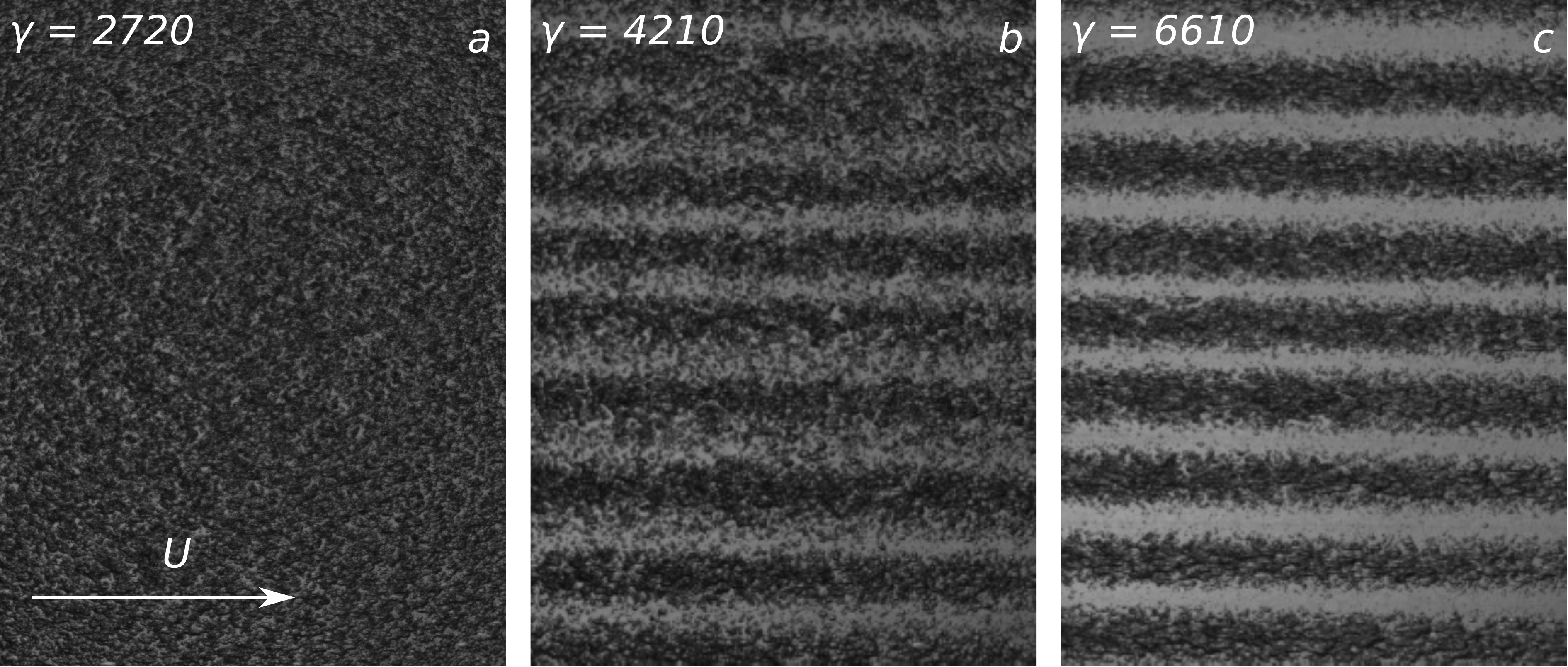}
  \caption{Example of vorticity banding in a plane parallel to the walls: droplet distribution at increasing time ($\gamma = \dot{\gamma}t$ is a non-dimensional measure of time) from left to right. Figure adapted from \citeauthor{Caserta2012}\citep{Caserta2012}.}
  \label{fig:caserta}
\end{figure}

The formation of banded structures in shear flows has been observed for different types of complex fluids \citep{Butler1999,Olmsted2008}. These structures can have different orientations depending on the flowing material: banding in the direction of the velocity gradient has been observed in worm-like micellar solutions as a consequence of a flow instability \citep{Spenley1993} whereas structures oriented in the vorticity direction and alternated in the flow direction have been reported for attractive emulsions \citep{Montesi2004}. \citeauthor{Kang2006}\cite{Kang2006} performed experiments of rodlike virus suspensions and observed vorticity banding in a limited range of shear rates which was explained in analogy to elastic instabilities of polymers (the Weissenberg effect): inhomogeneities in the flow induce a weak rotational flow in the gradient direction.\cite{Kang2008} Of interest here, vorticity banding has been observed in emulsions composed by a biphasic polymer blend flowing in the Newtonian regime \citep{Caserta2008,Caserta2012}. In these works, the authors perform experiments at different viscosity ratios (from 0.001 to 5.4) and shear rate (from 0.005 s\textsuperscript{-1} to 5 s\textsuperscript{-1}) in a range from dilute to moderate concentration of the disperse phase (the volume fraction ranging from 2.5\% to 20\%). They report the generation of droplet-reach and droplet-poor regions, regularly aligned in the direction of the flow and alternating in the vorticity direction, as illustrated in figure \ref{fig:caserta}. The vorticity banding phenomenon was observed only when the droplet phase showed a viscosity lower than the matrix phase (viscosity ratio < 1). The process, which leads after long time to a separation of the phases, has been associated with a decrease of the effective viscosity which is more evident for lower values of the viscosity ratio. \citeauthor{Caserta2012}\citep{Caserta2012} linked the formation of bands with the change of the concavity of the viscosity-volume fraction master curve. In this study we investigate the stability of banded structures and relate the vorticity banding in shear flows to the viscosity-concentration curve.

Emulsions are a biphasic liquid-liquid system in which the two fluids are partially or totally immiscible. Depending on the interaction force between the droplets, the emulsions can be considered repulsive or attractive. In the former case, the repulsive force between the drops is predominant, whereas in the latter case the system can produce flocculates \citep{Montesi2004}. The macroscopic properties of these systems strongly depend on their microstructure, mainly the droplet size and distribution. The two mechanisms that affect the disperse phase dynamics are the interface deformation and the collision rate. The deformation is due to the stress induced by the flow, counterbalanced by the interfacial stresses, which tend to reduce the droplet surface by keeping a spherical shape. The ratio of these two stresses is known as capillary number, Ca. When Ca exceeds a critical value, droplets do not reach a steady state shape under flow, but break in two or more fragments, each having a stable deformed shape \cite{Cristini2003}. On the other hand, in the case of non diluted emulsions, the flow can induce the collision of two or more droplets. During these interactions attractive forces between droplets can induce coalescence, provided the resulting larger drop is still stable under flow, \emph{i.e.} its capillary number does not exceeds the critical value. The interplay of these two mechanisms is of fundamental importance for describing the properties of liquid-liquid systems such as emulsions \citep{Caserta2005,Kostoglou2001}. These are important also in microfluidic devices to control the formation of droplets and manipulate their distribution, {\emph{e.g.}} T junctions or nozzles \citep{Teh2008,Sibillo2006}.

While the breakup involves mainly the interaction between one single drop and the external flow, the coalescence arises from the interaction of different drops, which complicates the dynamics of the system. If the Peclet number, defined as the ratio of the diffusion induced by the external flow and the molecular diffusion, is larger than 1 then the coalescence is flow-driven, which is the case under consideration in this study. Under this condition, it has been shown experimentally that coalescence decreases with increasing shear rate as well as with the particle-size ratio, due to changes in the trajectory of smaller droplets.\cite{Lyu2000,Lyu2002} The complex dynamics describing the behavior of two colliding droplets can be thought as the interplay of an external flow, responsible of the frequency, force and duration of collisions, and an internal flow (the drainage film between the two particles) which accounts for the deformation of the interface and, eventually, rupture and confluence. \cite{Chesters1991}. For spherical particles of equal size at low Reynolds number, it is possible to estimate a collision frequency per unit time and volume as $C = (2/3)\dot{\gamma}d^3n^2$, with $\dot{\gamma}$ the applied shear rate, $d$ the diameter of the particles and $n$ the number of particles per unit volume \cite{SmoluchowskiM.1917}. If the characteristic collision duration is larger than the drainage time, droplets will tend to coalesce, whereas the emulsion will behave as repulsive in the opposite case\cite{Guido1998}. From scaling analysis it is possible to approximate the drainage time in a head-on collision of two equal-size drops as $t_d \dot{\gamma} \approx Ca^m$, where $m = 4/3$ if the drainage film is assumed to be flat \citep{Chesters1991} or $m = 1$ for dimpled-film shape \citep{Frostad2013}. This estimation has also recently been corrected to account for the slip condition at the interface between polymers \citep{Ramachandran2016} which can give important differences mostly at low $Ca$. In a real scenario, the assumption of head-on collision is not always proper and the emulsion can have a polydisperse size distribution, which makes the previous estimate not fully reliable. The morphology of a liquid-liquid system under flow is a non trivial function of the flow intensity, depending on the entire flow history\cite{Minale1997,Minale1998}. Several experimental studies have been conducted on droplet collisions in shear flows in order to describe the size evolution and deformation \citep{Guido1998,Burkhart2001,Vananroye2006,Sibillo2006} and to investigate the effect of the wall confinement \citep{Chen2009a}.

Performing numerical simulations of droplet collisions and coalescence is a challenging problem due to the large separation of scales involved in the problem: from the external flow lengthscale, the gap between the two plates which can be order of mm, to the smallest scale given by the thickness of the fluid film between two drops, which can be order of nm \citep{Chesters1991}. Additionally, the process of band generation requires thousands or tens of thousands strain units \citep{Caserta2012} to fully develop, making the observation window very long. Fully resolved three-dimensional simulations of emulsions in shear flow, with same physical parameters as in experiments, are therefore extremely expensive. Numerical studies of emulsions at moderate concentration in literature have mostly been conducted with methods that do not allow droplets to coalesce \citep{Loewenberg1996,Srivastava2016} whereas simulations which resolve the liquid films are mostly in the dilute regime \citep{Shardt2013,DeBruyn2013}. 

In this work we present a numerical investigation of emulsions in shear flow with volume fraction $\phi$ of the disperse phase ranging from 5\% to 20\% and viscosity ratio $\lambda$ from 0.01 to 10, defined as the ratio of the disperse phase viscosity over the outer fluid viscosity. To avoid to simulate the long process of bands generation, occurring over thousands of shear units, the initial condition of our simulations is a distribution of droplets already in the forms of bands, whose stability is investigated for different coalescence efficiency. We aim to first reproduce the experimental observation in \citeauthor{Caserta2012}\citep{Caserta2012} and to explain the vorticity banding process by the effect of the coalescence on the droplet distribution and on the rheological behavior of the system.

\section{Numerical Method and Setup}

We simulate emulsions at moderate concentrations in shear flows at low Reynolds number. The multiphase flow is governed by the incompressible Navier-Stokes equations
\begin{subequations}
  \begin{align}
    \label{eqn:navier-stokes}
    \frac{\partial u_i}{\partial x_i} &= 0, \\
    \rho\left(\frac{\partial u_i}{\partial t} + u_j\frac{\partial u_i}{\partial x_j}\right) &= -\frac{\partial p}{\partial x_i} + \frac{\partial}{\partial x_j}\left(2\mu D_{ij}\right) + \sigma \kappa \delta_s n_i.
  \end{align}
\end{subequations}
where $u_i$, with $i = 1,2,3$, are the velocity components in the three Cartesian coordinates $x_1, x_2$ and $x_3$, $p$ the pressure field, $\rho$ and $\mu$ the local density and viscosity, $\mathbf{D}$ the rate of deformation tensor $D_{ij} = \left(\partial u_i / \partial x_j + \partial u_j / \partial x_i\right)/2$, $\sigma$ the interfacial tension coefficient, $\kappa$ the curvature of the interface, $n_i$ the $i-th$ component of the unit normal vector $\mathbf{n}$ to the interface and $\delta_s$ the Dirac delta function which express that the interfacial tension force acts only at the interface between the two fluids. To track in time the position of the interface we employ a Volume of Fluid (VoF) technique based on the multi-dimensional tangent of hyperbola interface capturing (MTHINC) method \citep{Ii2012}. To identify the two fluids we introduce a VoF function $\mathcal{H}(\mathbf{x},t)$ defined as the cell-average value of the volume fraction of one fluid in the other. The VoF function is advected by the flow field as
\begin{equation}
  \label{eqn:vof}
  \frac{\partial \mathcal{H}}{\partial t} + u_j\frac{\mathcal{H}}{\partial x_j} = 0.
\end{equation}
The material properties of the two fluids are linked to the VoF function $\mathcal{H}$ as follow
\begin{equation}
  \begin{aligned}
    \rho(\mathbf{x},t) = \rho_1\mathcal{H}(\mathbf{x},t) + \rho_0(1-\mathcal{H}(\mathbf{x},t)) \\
    \mu(\mathbf{x},t) = \mu_1\mathcal{H}(\mathbf{x},t) + \mu_0(1-\mathcal{H}(\mathbf{x},t))
  \end{aligned}
\end{equation}
where the subscript 1 stands for the disperse phase, the subscript 0 for the carrier fluid and $\mathcal{H}$ is equal to 1 in the disperse phase and 0 in the carrier fluid. Finally the surface tension force is approximated using the Continuum Surface Force (CSF) approach \citep{Brackbill1992}
\begin{equation}
  \sigma \kappa \delta_s n_i = \sigma \kappa \frac{\partial \mathcal{H}}{\partial x_i}.
\end{equation}
The solver uses the second-order centered finite difference scheme for the spatial discretization and a fractional step algorithm for the time marching with a fast FFT solver for the resulting Poisson pressure equation. For the temporal discretization a second order Adam-Bashfort scheme is used. See \citeauthor{Rosti2018b}\cite{Rosti2018b} for a detailed description and validation of the code employed in this work.

\begin{figure}[h]
  \centering
  \includegraphics[width=\linewidth]{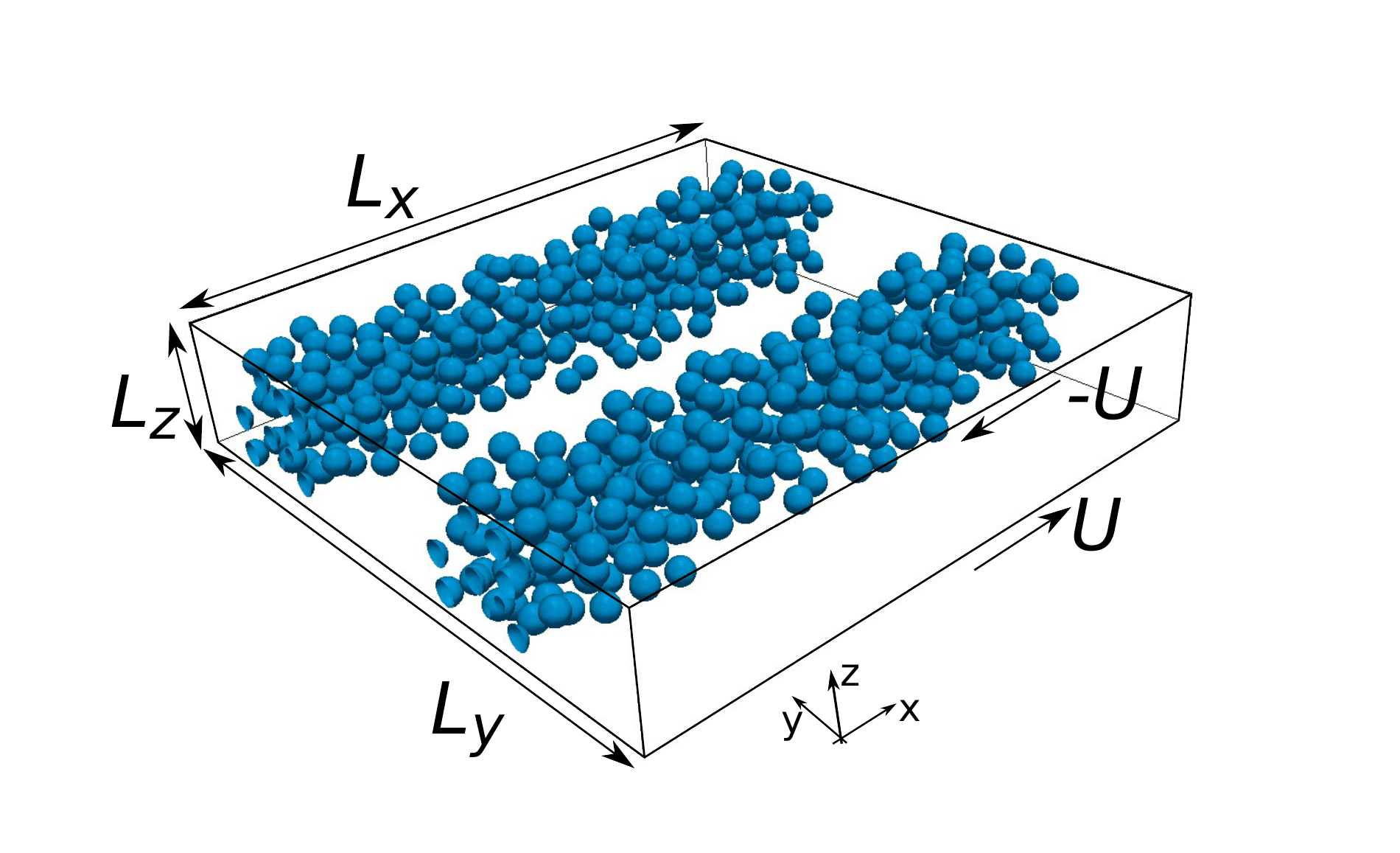}
  \caption{Sketch of the computational geometry and of the initial configuration.\label{fig:init}}
\end{figure}

\begin{figure}
  \centering
  \includegraphics[width=\columnwidth]{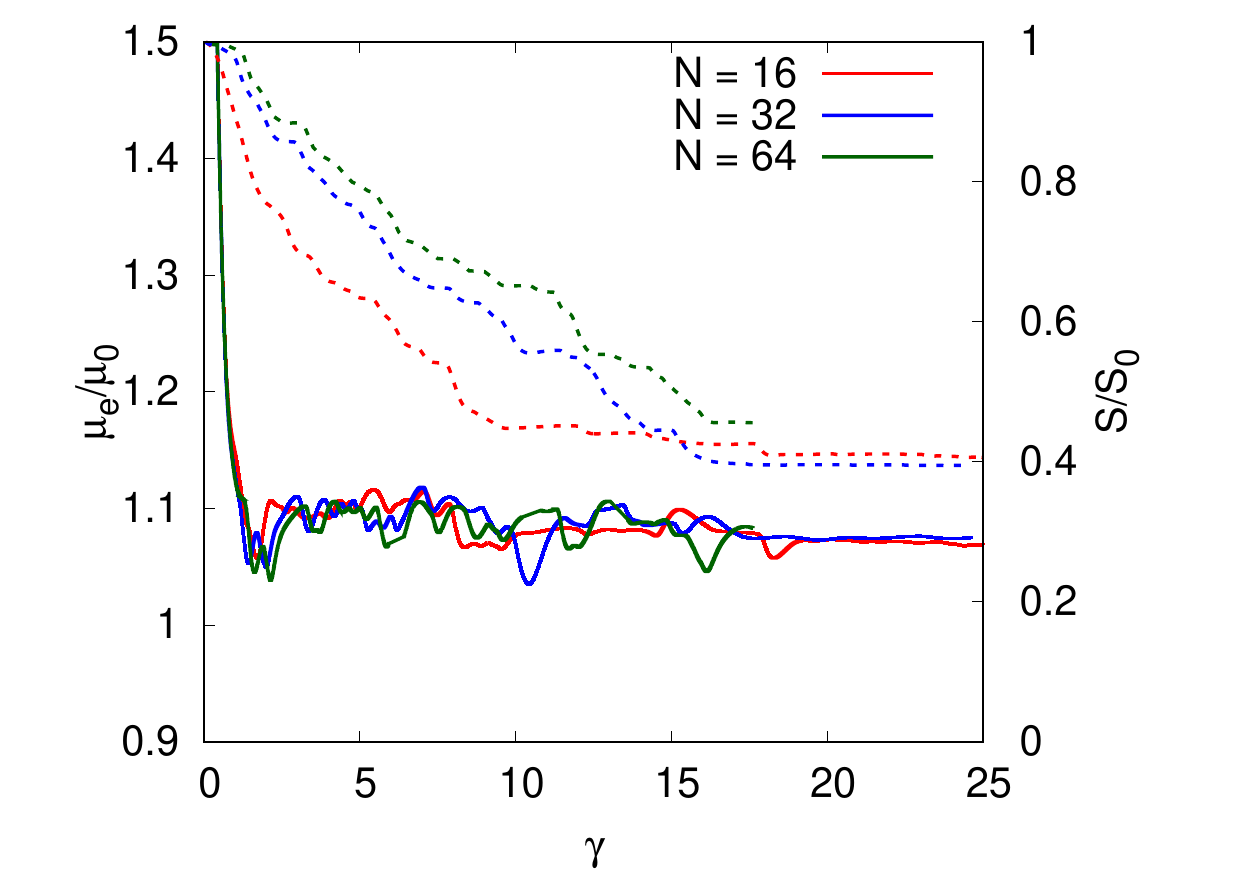}
  \caption{Time history of the effective viscosity (solid lines) and normalized surface (dashed lines) for three different grid resolution: $N = 16$ (red); $N = 32$ (blue), $N = 64$ (green), with $N$ the number of grid points  per initial radius. These results refer to the simulation in the small domain.\label{fig:convergence}}
\end{figure}

We know from previous experimental studies\cite{Caserta2012} that, under specific conditions, sheared emulsions form vorticity aligned banded structures. The process of formation of bands from an isotropic emulsion needs more than 1000 strain units (the strain $\gamma = \dot{\gamma}t$ is a measure of the overall deformation imposed to the sample, and can be considered as a non-dimensional measure of time under flow). Running a numerical simulation of this process for the entire time required to complete its dynamic is not feasible with the actual computational limits. For this reason, we decided to initialize the disperse phase with the already-formed banded structures and verify in which conditions they are stable and in which they will diffuse. The characteristic width of the bands is of the order of the gap between the plates $\delta$\citep{Caserta2012}, which clearly highlights the effect of the confinement on the phenomenon under study. We initialize a random distribution of droplets, of equal radius $r$, confined in two bands of width order $\delta$. It is worth noticing that in the case of morphological hysteresis the steady state droplet size distribution is not only function of the applied shear rate but depends also on the initial distribution.\cite{Minale1997} Thus, the focus of this study is not on the droplet morphology at steady state but rather on the stability of banded structures. To characterize the rheological behavior of the emulsion we also compute the effective viscosity in a small domain starting with an homogeneous droplet distribution and preventing the banding in the vorticity direction by increasing the lateral confinement. By doing so, we are able to compute the constitutive curve of the system. In the following we will refer to the two cases as large domain (LD) and small domain (SD).

\begin{figure*}[h]
  \centering
  \begin{subfigure}{0.42\textwidth}
    \includegraphics[width=0.9\textwidth]{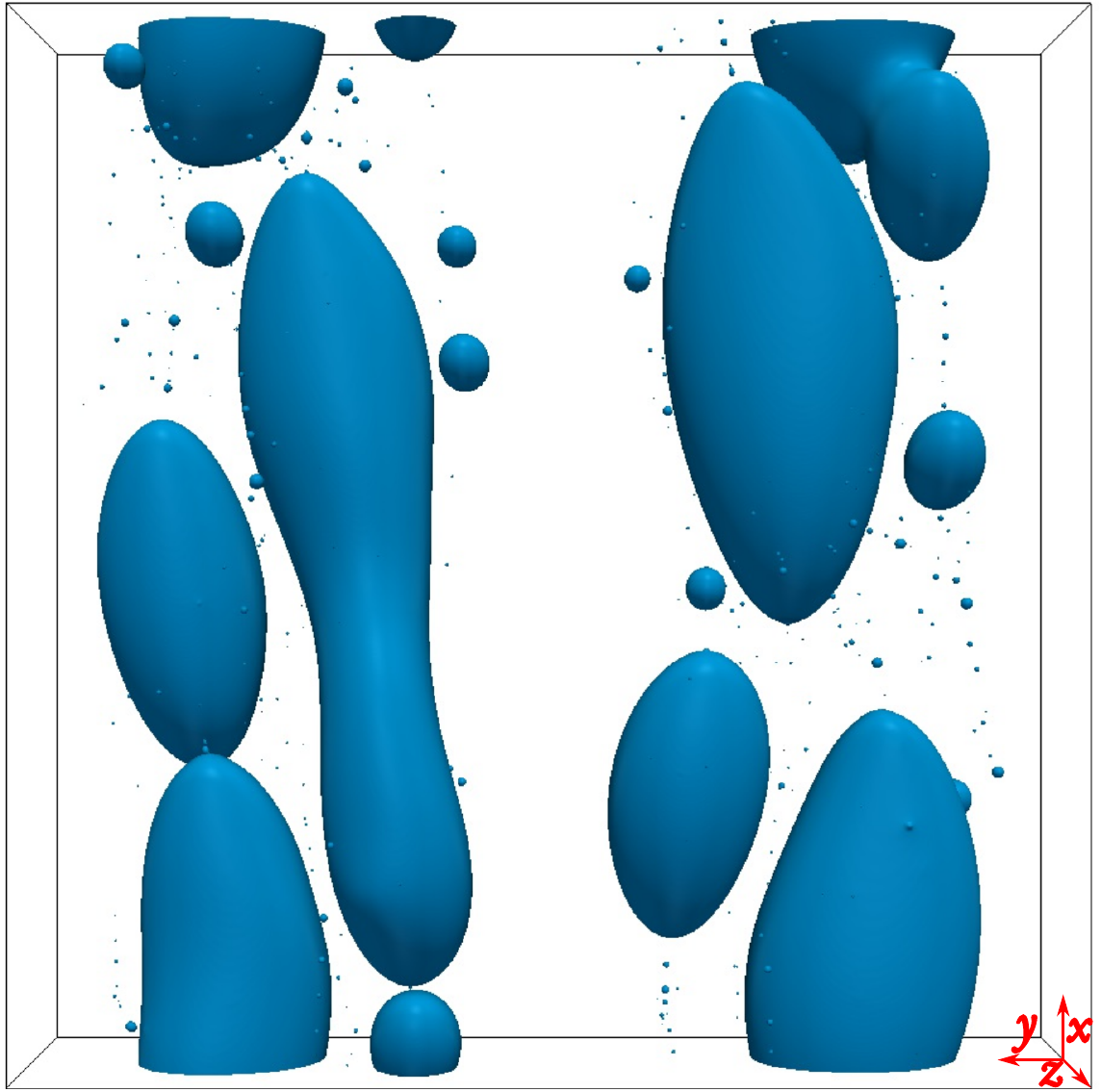}
  \end{subfigure}
  \begin{subfigure}{0.56\textwidth}
    \includegraphics[width=0.9\textwidth]{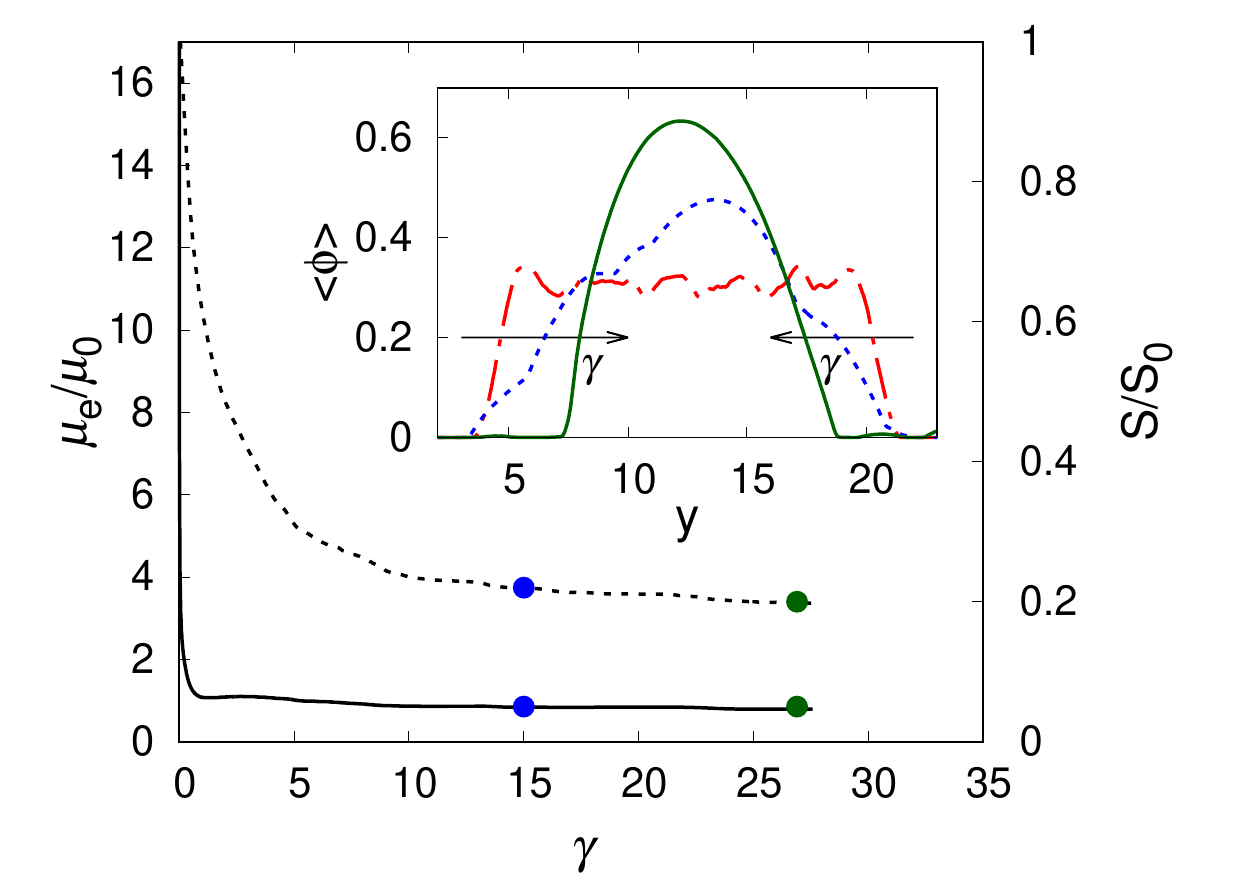}
  \end{subfigure}
  \caption{(Left) Droplets distribution at strain unit $\gamma = 15$ for the simulations with 20\% volume fraction. (Right) Time history of the effective viscosity $\mu_e$ normalized with the outer fluid viscosity $\mu_0$ for the simulation with $\phi = 20\% $ (solid black line); on the right vertical axis, normalized total surface for the simulation with $\phi = 20\%$ (black dotted line). The inset shows the average volume fraction $<\phi>$ in the vorticity direction ($y$) at three different instants for the case with volume fraction 20\%: initial distribution (dashed-dot red line); distribution at $\gamma = 15$ (dotted blue line); distribution at $\gamma = 27$ (solid green line).}
  \label{fig:bande_merging}
\end{figure*}

A sketch of the computational domain and the initial distribution of the droplet is reported in figure \ref{fig:init}. The domain is periodic in the $x$ and $y$ directions (velocity and vorticity directions) and wall bounded in the $z$ direction (velocity gradient direction). The large domain box has size $L_x = L_y = 10\delta$ and $L_z = \delta$ whereas the small domain has size $1.6\delta$ in the $x$ and $y$ direction and same size in the $z$ direction, with $\delta = 10r$. The top and bottom walls move with opposite velocities $\pm U$ such that the applied shear rate is equal to $\dot{\gamma} = 2U / \delta$, chosen to ensure that the droplet Reynolds number $Re = \dot{\gamma}r^2/\nu_0$ is equal to 0.1, being $\nu_0$ the kinematic viscosity of the outer fluid. The interfacial tension coefficient $\sigma$ is chosen to provide a capillary number, based on the initial radius $Ca = \mu_0 \dot{\gamma} r/\sigma$, equal to 0.1, matching the experiments in \citeauthor{Caserta2012}\citep{Caserta2012}. In the experiments the banding has been observed only for viscosity ratios smaller than one \citep{Caserta2012}; additionally the time required to reach a stable and steady configuration is shorter for smaller $\lambda$ \citep{Caserta2008}. For this reason we simulate banding emulsions only with a viscosity ratio of 0.01. The coalescence probability estimated as in \citeauthor{Chesters1991}\citep{Chesters1991} with the chosen set of parameters is equal to 0.9, thus suggesting that almost every collision will results in droplet coalescence. Additionally, the effect of breakup is secondary compared to coalescence as observed experimentally \cite{Caserta2012} and also verified \emph{a posteriori} in our simulations. All the simulations have been performed with a resolution of 32 grid points per droplet initial diameter. We have verified that this resolution is enough to properly describe both the transient and steady state behavior of the emulsions, as shown in figure \ref{fig:convergence}. In particular, the effective viscosity has already converged for the coarser gird (16 points per initial radii), whereas the coalescence at low resolution is slightly faster (10\% difference) but the steady state value is the same. From the simulation with 64 points (shorter in time for computational reasons) we verified that the adopted resolution ($N = 32$) is also enough to capture the transient dynamics with a maximum error of less then 8\% in the total surface. In this study, the effective viscosity $\mu_e$ is always computed as the ratio between the wall shear stress (\emph{i.e.} the time and space average of the derivative in the $z$ direction of the horizontal velocity $u$ at the walls multiplied by the outer fluid viscosity) and the applied shear rate 
\begin{equation}
  \mu_e = \frac{ \mu_0 <<\frac{\partial u}{\partial z}|_w>>}{\dot{\gamma}}
\end{equation}
where the symbol << >> indicates time and space average in the homogeneous directions.

\section{Results and discussion}

In absence of any potential, for example gravitational or electric field, the dynamics of the disperse phase is governed only by the external flow. As the simulations start, the shear is transferred from the plates to the interior of the domain, the drops start to deform, align in the direction of the shear and eventually collide. As soon as two pieces of interface are in the same computational cell they will merge leading to the coalescence of the drops. In this case every collision leads to coalescence, which implies an overestimation of the coalescence efficiency of the system. This is a well-known issue of interface capturing methods as the VoF solver employed in this study. From a physical point of view, this is equivalent to having a system with drainage time tending to zero, or coalescence efficiency tending to unity. The condition of unitary coalescence efficiency results in an extremely fast dynamic evolution of the droplet size under flow, leading to the formation of very large droplets. As droplet size grows the capillary number approaches the critical value, so an equilibrium between droplet coalescence and breakup could be the expected steady-sate. However in confined conditions, as in our study, large droplets can be stabilized by the presence of walls \cite{Sibillo2006}. This is in agreement with what was observed in the experiments where, at very large strain values, and in the case of high coalescence efficiency, extremely large drops are visible. \cite{Caserta2008,Caserta2012}. Analogous structures have been also reported by \cite{Migler2001}. We consider the simulations to be steady when the effective viscosity vary less than 3-4\% in a period of about 10 strain units.

We display in figure \ref{fig:bande_merging} the instantaneous distribution of the droplets (left panel) after $15$ strain unit for volume fraction equal to 20\%. Droplets inside the bands, where collisions are more frequent, start to coalesce and create bigger structures elongated in the direction of the shear. The distribution of the average volume fraction $<\phi>$ in the vorticity direction ($y$), computed by averaging in the $x$ and $z$-directions the local volume fraction $\phi$, is reported in the inset of the right panel for three different instants, also marked in the time history of the effective viscosity and total surface in the same figure. As a consequence of the coalescence, the average concentration becomes more peaked and confined in space, as visible by comparing the initial condition (red dash-dotted line) to the blue and green curves in the inset. In other words, in these conditions we observe that the initial banded structures are stable, \emph{i.e.} they remain localized in their initial position and do not diffuse along the vorticity direction. Note also that the initial width of the band was about $1.5\delta$ but at steady state it becomes $\delta$. To verify the independence of this result from the initial configuration we considered an additional case with bands of local volume fraction 30\%  and with a non-zero volume fraction, about 5\%, between the bands so that the mean volume fraction in the domain is 15\%, and also for this configuration the initial banded structure is stable (not shown here).

\begin{figure}[h]
  \centering
  \includegraphics[width=\columnwidth]{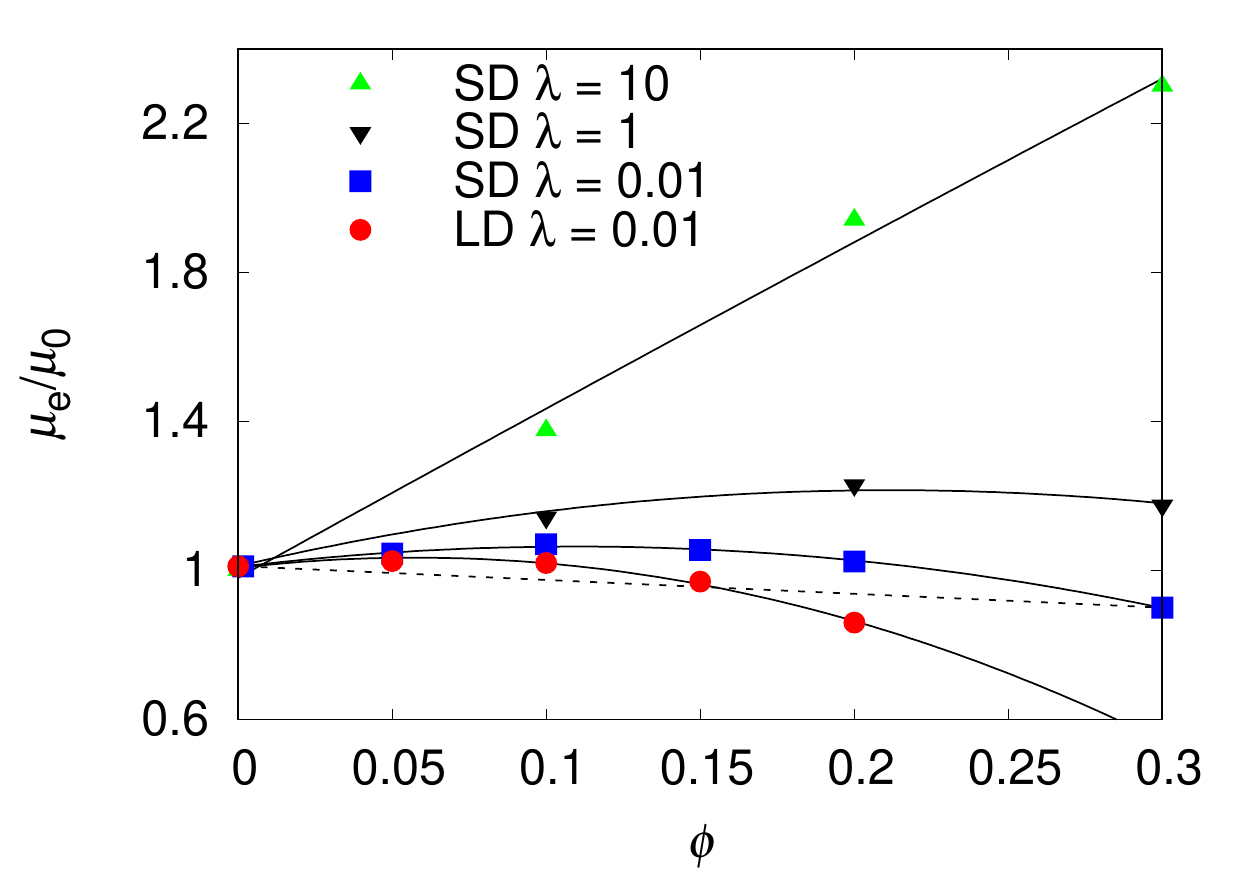}
  \caption{Comparison of the effective viscosity $\mu$ normalized with the outer fluid viscosity $\mu_0$ between simulations in the large domain (LD) and $\lambda = 0.01$ (red dots) and simulation in the small domain (SD): $\lambda = 0.01$ (blue square), $\lambda = 1$ (black triangle), $\lambda = 10$ (green triangle). The black lines are polynomial fit to the data. The dashed line represent the level rule between two phases, one given by droplet rich areas at $\approx$ 30\% $\phi$ and one by droplet poor area at $\approx$ 0\% $\phi$, as explained in the text.}
  \label{fig:mue_coalescence}
\end{figure}

\begin{figure}[h]
  \centering
  \begin{subfigure}{0.45\textwidth}
    \includegraphics[width=0.9\textwidth]{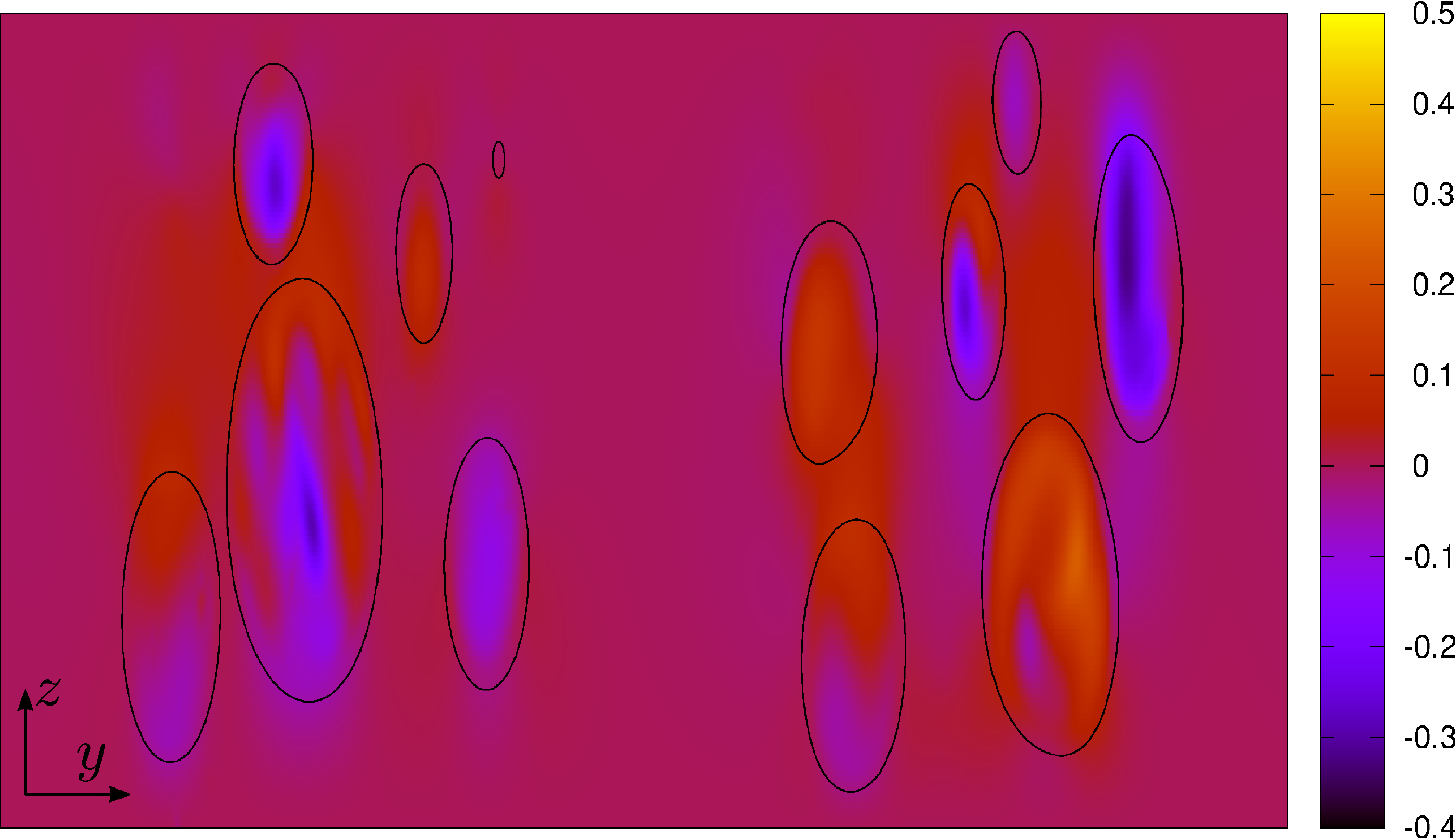}
  \end{subfigure}
  \begin{subfigure}{0.45\textwidth}
    \includegraphics[width=0.9\textwidth]{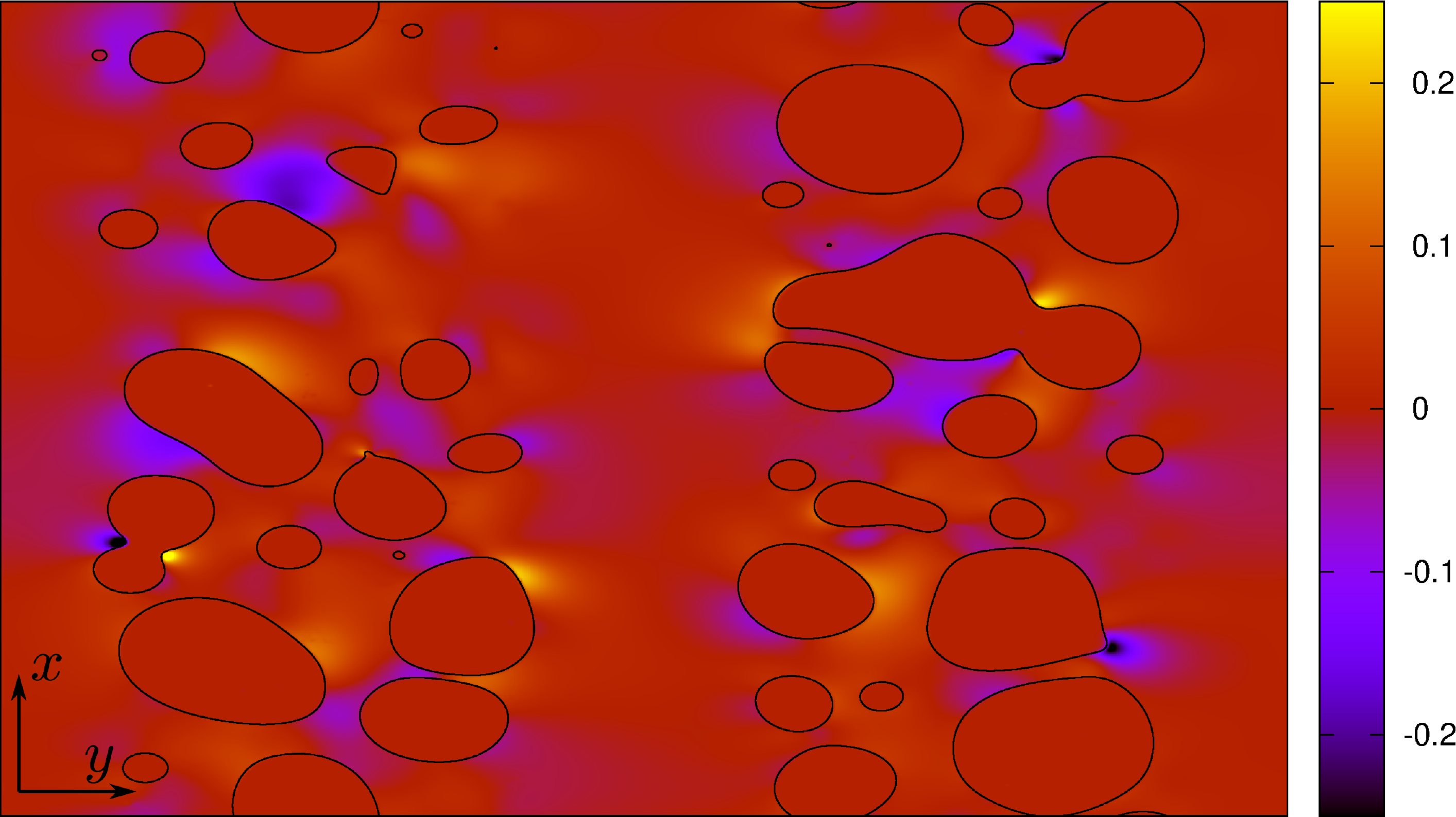}
  \end{subfigure}
  \caption{(Top panel) Snapshot of the velocity component in the gradient direction ($z$) showing the migration of the droplets to the center of the domain. (Bottom panel) Snapshot of the velocity component in the vorticity direction ($y$). For both panels $\gamma = 5$.}
  \label{fig:mechanism}
\end{figure}

To understand why banded structures are stable in this condition it can be helpful to consider the shape of the constitutive curve of an emulsion. Conversely to rigid\citep{Guazzelli2018} and deformable particles\citep{Rosti2018a}, where the effective viscosity always increases with the volume fraction, emulsions can exhibit a constitutive curve of the effective viscosity vs the volume fraction with a negative curvature when the viscosity ratio is lower than unity \cite{Caserta2012,devita2019}. As previously discussed, we computed the effective viscosity of an homogeneous emulsion in a smaller domain in the homogeneous directions and same vertical distance between the walls $\delta$. This smaller domain prevents the separation of phases hence the system can be seen as representative of a homogeneous distribution with a certain average volume fraction. It is worth noticing that, the viscosity computed in the small domain (SD) represents a constitutive curve of the emulsion, being the distribution homogeneous, whereas in the large domain (LD) we measure the flow curve of the system. In figure \ref{fig:mue_coalescence} we compare the effective viscosity of the two systems, the large domain with bands (morphologically described in figure \ref{fig:bande_merging}) and the homogeneous small domain with no bands. The results illustrate that for viscosity ratio smaller than 1 the constitutive curve has negative curvature; this effect reduces increasing $\lambda$ so that the curvature becomes positive for viscosity ratio greater than 1. By comparing the effective viscosity between the simulations in the two different domain we also confirm the prediction of the experiments showing that the presence of banded structures effectively reduces the viscosity of the system. This effect is a direct consequence of the curvature of the constitutive curve: for instance, an emulsion with average volume fraction $\phi = 15\%$ in a small domain will have an effective viscosity given by the blue square in figure \ref{fig:mue_coalescence}. If we consider the same volume fraction but in a larger domain, which is able to fit the banded structures, the phases will tend to separate producing droplet-rich areas (in the example at 30\% volume fraction) and droplet poor-areas (approximately 0\% volume fraction). The effective viscosity of this system, given by the red circle in correspondence of $\phi = 15\%$, lies approximately on the dashed line connecting the values of the viscosity at 0\% and 30\% for the flow without bands. In other words, it is possible to apply the level rule by considering the droplet rich and droplet poor regions as two different phases with volume fraction $\phi_1 = 0.3$, $\phi_2 = 0$ and effective viscosity $\mu_1 = 0.9$ and $\mu_2 = 1$, respectively. If $\xi$ is the volume occupied by the bands with respect to the whole domain (approximately 0.5), the average volume fraction is given by $\phi = \xi \phi_1 + (1-\xi) \phi_2 = 0.15$. Similarly, for the effective viscosity we have $\mu = \xi \mu_1 + (1-\xi) \mu_2 = 0.95$ which is approximately the value given by the red dots for $\phi = 0.15$. This implies that, due to the shape of the constitutive curve, if the domain is large enough the phases will separate because the final state is energetically more convenient, being associated to a lower effective viscosity. Additionally, since the two curves diverge, the decrease in effective viscosity given by the phase separation is more significant for higher volume fractions. For small $\phi$, instead, since the average distance between droplets increases collision frequency tends to zero and the vorticity banding would require extremely long time to form and is unlikely to be seen. This can also be explained considering the level rule and observing that for small volume fraction the difference in the effective viscosity between the banded and non-banded distribution becomes negligible, hence the curvature of the master curve becomes almost zero. Note also that, by increasing the viscosity ratio, the curvature of the constitute curve becomes positive and for $\lambda >1$ the banding is not anymore energetically convenient. These results are in agreement with the experimental observations by \citeauthor{Caserta2012}\citep{Caserta2012}. The reason why at large viscosity ratio there is no banding is that the coalescence is greatly favored by a low disperse phase viscosity and reduces significantly for $\lambda > 1$\citep{Chesters1991}; as we will see in the next section, coalescence is the key ingredient for the banding.

\begin{figure*}[h]
  \centering
  \begin{subfigure}{0.45\textwidth}
    \includegraphics[width=0.9\textwidth]{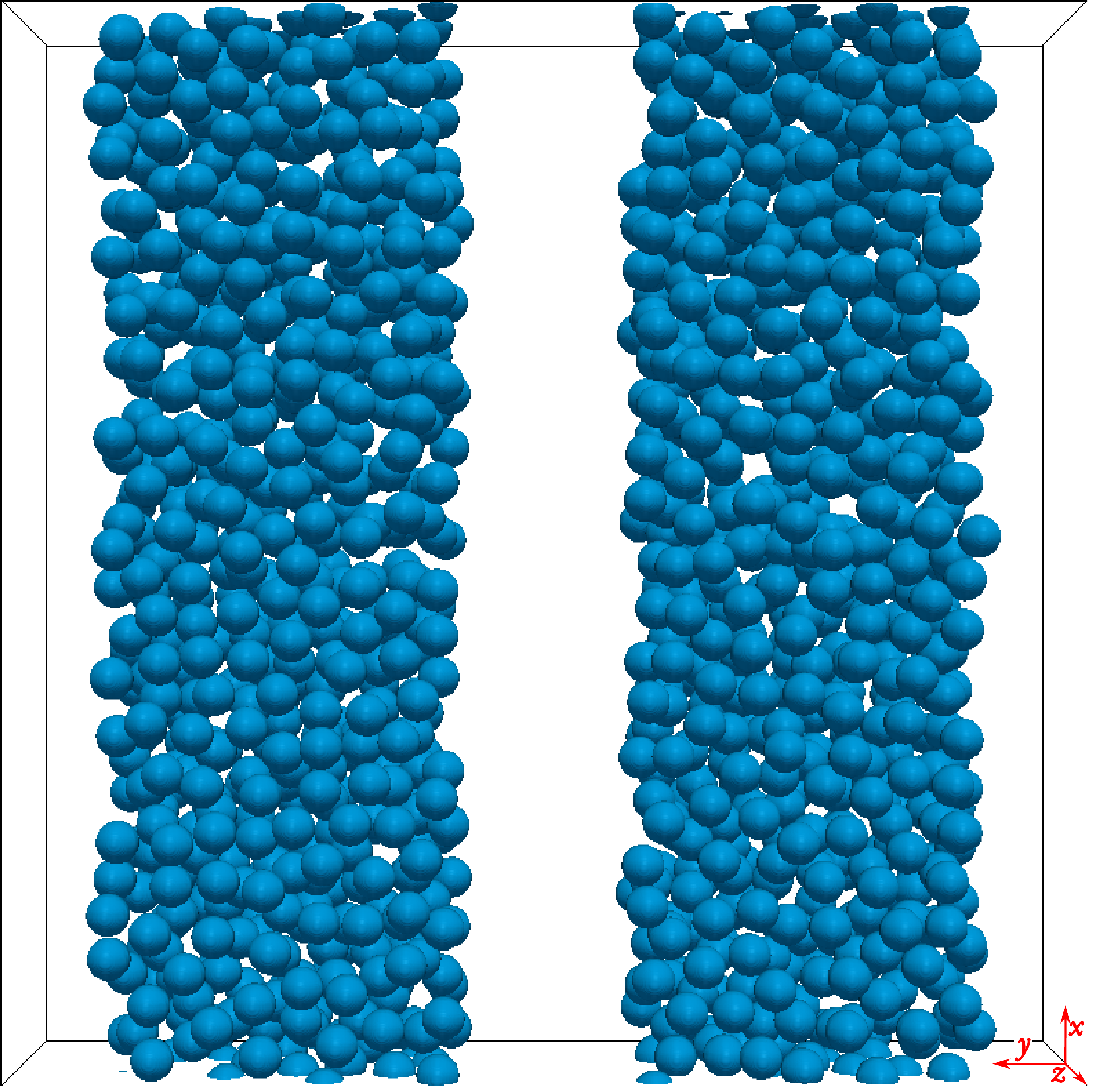}
  \end{subfigure}
  \begin{subfigure}{0.45\textwidth}
    \includegraphics[width=0.9\textwidth]{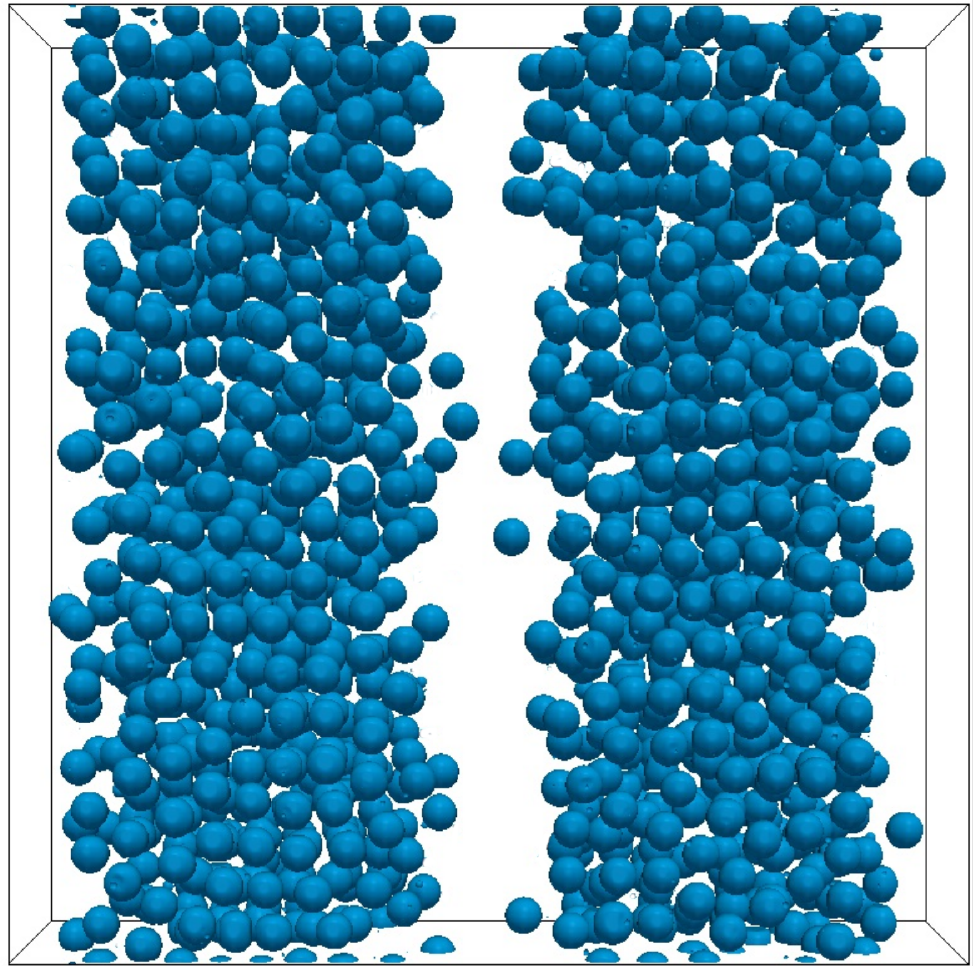}
  \end{subfigure}
  \caption{Drops distribution for the simulations with collision force (i.e. coalescence efficiency tending to zero) for the case with 20\% average volume fraction: $\gamma = 0$ (left), $\gamma = 105$ (right).}
  \label{fig:bande_collision}
\end{figure*}

\begin{figure}[h]
  \centering
  \includegraphics[width=\columnwidth]{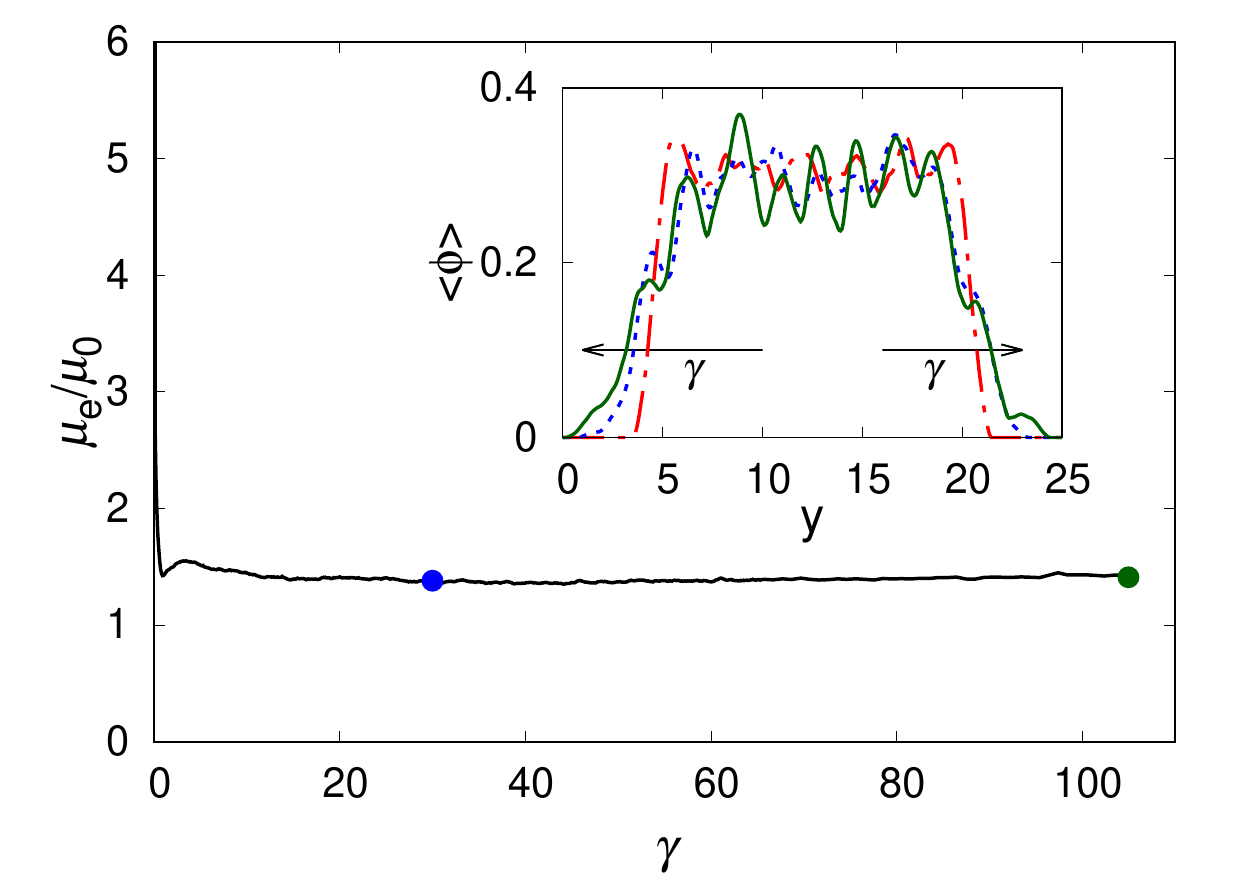}
  \caption{Time history of the effective viscosity $\mu_e$ normalized with the outer fluid viscosity $\mu_0$ for the simulation with $\phi = 20\%$ and collision force. The inset shows the average volume fraction $<\phi>$ in the vorticity direction ($y$) at three different instants: initial distribution (dashed-dot red line); distribution at $\gamma = 30$ (dotted blue line); distribution at $\gamma = 105$ (solid green line).}
  \label{fig:distribution_collision}
\end{figure}

Before proceeding to the next section we want to suggest an analogy between our system and the mechanism proposed in Ref \citenum{Kang2008}. \citeauthor{Kang2008}\cite{Kang2008} observed vorticity banding in suspensions of rodlike viruses for a specific range of the applied shear rate. They proposed a mechanism for the generation of these bands inspired by the elastic instability of polymers: the presence of inhomogeneities in the flow induces normal stresses in the gradient direction which, in turn, produce a weak flow causing the phase separation in banded structures. In our system inhomogeneities correspond to droplets of different size, whereas the flow in the gradient direction ($z$) is produced by the migration of larger droplets towards the center of the gap (see figure \ref{fig:mechanism} top panel). The latter effect has been demonstrated experimentally \cite{Hudson2003,Caserta2005} and previous simulations have also shown that it is related to the reduction of the emulsion effective viscosity \cite{devita2019}. Once droplets become larger and migrate towards the center of the gap, they also move slower in the flow direction than smaller droplets. This induces a gradient of the flow velocity in the vorticity direction, associated with a weak flow in the vorticity direction (see figure \ref{fig:mechanism} bottom panel).

\subsection{Effect of drainage time}

As mentioned before, our simulations represent the limit of drainage time tending to zero or, equivalently, of unitary coalescence efficiency. What will happen if we consider a system with drainage time tending to infinity? In this case the collision time is always faster than the drainage time and coalescence never occurs. To reproduce this limiting case in our simulations, we introduce an Eulerian subgrid force\citep{Bolotnov2011} in the momentum equation that depends on the signed distance $\psi$ (usually referred to as level-set function) from a droplet interface
\begin{equation}
  \mathbf{F}_c = \mu_0 U r \left( \frac{a}{\psi} + \frac{b}{\psi^2} \right)  \mathbf{n}
\end{equation}
where $\mu_0$ is the outer fluid viscosity, $r$ the initial radius of the droplets, $U$ the wall velocity, $\mathbf{n}$ the normal to the interface and $a$ and $b$ are two coefficients. Every timestep the distance function $\psi$ is reconstructed from the VoF field solving the following redistancing equation
\begin{equation}
  \label{eqn:redistance}
  \frac{\partial \psi}{\partial \tau} + S\left(\psi_0\right)\left(|\nabla \psi|-1\right) = 0
\end{equation}
with $S\left(\psi_0\right)$ the sign function, $\tau$ an artificial time and the initial distribution $\psi_0$ given by $\psi_0 = (2\mathcal{H}-1)0.75\Delta$, being $\Delta$ the grid spacing \citep{Albadawi2013}. We solve equation \eqref{eqn:redistance} using the second order algorithm of \citeauthor{Russo2000}\cite{Russo2000} with a timestep $\Delta \tau = 0.1\Delta$. In the implementation, we tag every initial droplet with an index and whenever two different droplets are closer than three grid points we apply this repulsive force. By changing the magnitude of the force we can delay or totally prohibit coalescence, therefore this force can be thought of as the consequence of the additions of surfactants in the emulsion. In this study, the two coefficients $a$ and $b$ have been chosen to provide the smallest force able to completely prevent the coalescence of the droplets, which correspond to $a = 55$ and $b = 3.5$. See \citeauthor{devita2019}\cite{devita2019} for a detailed description of the algorithm and the effect of this collision force.

\begin{figure}[h]
  \centering
  \includegraphics[width=\columnwidth]{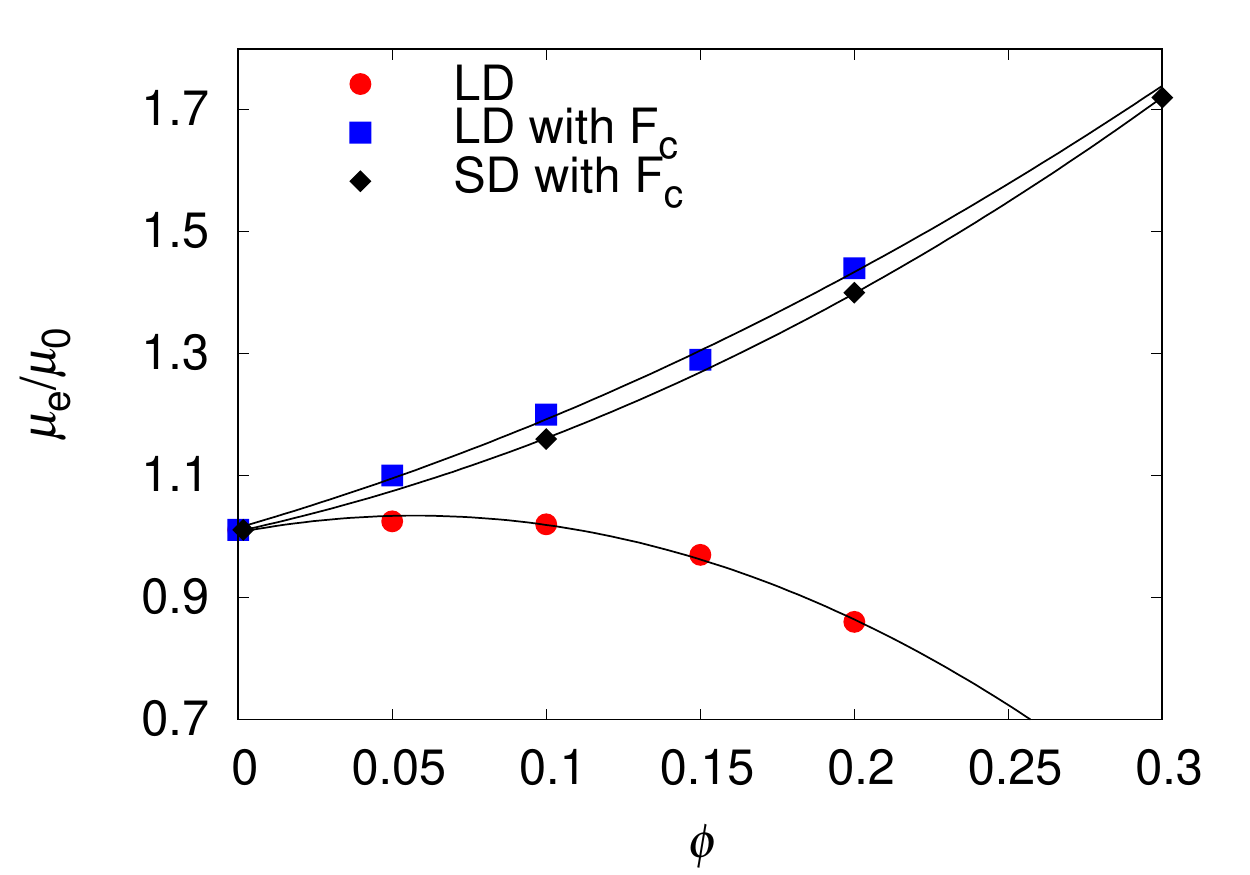}
  \caption{Comparison of the effective viscosity $\mu_e$ normalized with the outer fluid viscosity $\mu_0$ for the simulations without the collision force (red circle) and with collision force (blue square). The values for the cases with the collision force are taken at $\gamma = 25$. Black diamonds correspond to simulations in a small domain, as in figure \ref{fig:mue_coalescence}, starting with initial homogeneous distribution.}
  \label{fig:mue_collision}
\end{figure}

The evolution of the droplet configuration for the cases with collision force is reported in figure \ref{fig:bande_collision}. Unlike what observed in the previous cases, the initial banded structures are unstable in absence of coalescence and start to diffuse in the vorticity direction. Looking at the time evolution of the average volume fraction distribution $<\phi>$ (inset of figure \ref{fig:distribution_collision}), we notice that the bands become larger, the peaks reduce and the distribution tends to diffuse in the $y$-direction. If we now plot the effective viscosity vs the volume fraction for the cases with collision force we find that the effective viscosity grows with the volume fraction and that there is a change in the sign of the curvature (see figure \ref{fig:mue_collision}). It is worth noticing that when collision forces are active the effective viscosity obtained in large domains (blue squares) or for small domain where vorticity diffusion is inhibited (black diamonds), is approximately the same, as reported in the figure \ref{fig:mue_collision}. The small difference between the two curves with collision force is due to the remaining trace of the original bands in the simulations in the large domain, which diffuse very slowly. When the coalescence is prevented, the banded structures are not energetically convenient because the banded configuration exhibit an higher effective viscosity. This suggests that the initial banded structures, when coalescence is prohibited, tend to distribute homogeneously inside the domain.

When the collision force is applied to prevent the coalescence, the system behaves similarly to a suspension of deformable particles \citep{Rosti2018a,Rosti2018,Matsunaga2015}. \citeauthor{Rosti2018}\cite{Rosti2018} showed that the effective viscosity of a suspension of deformable particles can be estimated with the Eilers formula, valid for rigid spheres, by computing an effective volume fraction based on the mean deformation of the particles. Suspensions of deformable particles and droplet when they cannot merge have therefore a constitutive curve with a positive curvature. This implies that banded structures are not associated to a minimum of viscosity and thus are unstable; applying the same level rule as done in the case with merging will give a larger value of the effective viscosity for the banded case. The mechanism of coalescence allows emulsions to reduce the total surface of the system and thus to reduce the viscous dissipation associated to the flow. This also explains why the banding was not observed in experiments for viscosity ratio greater than unity \citep{Caserta2012}: for $\lambda > 1$ the coalescence efficiency reduces \citep{Chesters1991}.

\subsection{Shear stress budget}

\begin{figure}
  \centering
  \includegraphics[width=\columnwidth]{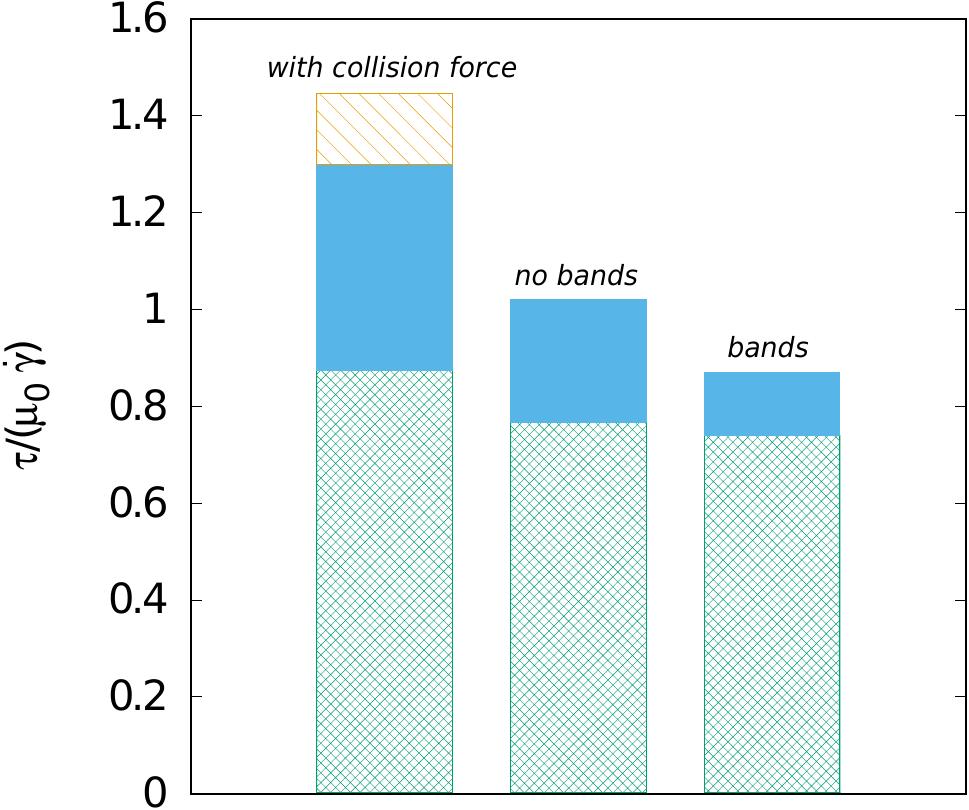}
  \caption{Shear stress budget for the case with collision force (left column), the small domain with no bands(middle column) and the large domain with bands (right column): outer viscous stress (green dense net); surface tension (solid blue), collision force (orange oblique lines). The contribution to the viscous stress in the inner fluid is about 100 times smaller than the others and is not visible on this scale.}
  \label{fig:histogram}
\end{figure}

To better understand this aspect, we compute the contribution of each term of the momentum equation to the shear stress. To this end, we consider the $x$-component of the momentum equation, average it in the homogeneous directions ($x$ and $y$) and integrate in the $z$-direction \citep{Rosti2018b}:
\begin{equation}
  \tau_{xz} = \tau_{xz}^{\mu} + \tau_{xz}^{\sigma} + \tau_{xz}^{c}  
\end{equation}
where the first term is associated to the viscous dissipation, the second term to the interface tensions and the last term to the collision force. To account for the interface tension we use the Continuum Surface Force approach \cite{Brackbill1992} in which the interface tension is expressed as $F_{\sigma} = \sigma \kappa \nabla \phi$, where $\kappa$ is the local curvature. The shear stress due to interfacial tension is computed as
\begin{equation*}
  \tau_{xz}^{\sigma}(z) = \int_z <F_{\sigma,x}>\,dz;
\end{equation*}
in the same way we compute the contribution to the stress due to the presence of the collision force as
\begin{equation*}
  \tau_{xz}^c(z) = \int_z <F_{c,x}>\,dz.
\end{equation*}
See \citeauthor{devita2019}\cite{devita2019} for a full derivation. The average values of all the contributions are reported in figure \ref{fig:histogram} divided by the Newtonian shear stress so that the sum of the components gives the suspension effective viscosity. The stress budget clearly shows that when the coalescence is prevented, most of the increase in the effective viscosity is due to the interface tension term. The stress associated to the deformation is proportional to the surface area which is higher in the cases with collision force. This confirms that the coalescence is the mechanism that allows emulsions to reduce their effective viscosity. From the graph, we can also observe that, with merging, the additional reduction of the effective viscosity in presence of shear banding is mostly related to a further reduction of the interfacial tension term, with a small contribution associated to the viscous dissipation. Indeed, for the simulations with bands we observe that the droplets tend to approach the walls more than for the case without bands, which can explain the small reduction of the viscous stress. The reduction of the total area in the simulation with bands is about 8\% higher which confirm the reduction of the interface tension stress.

\subsection{Bands diffusion}

\begin{figure}
  \centering
 \begin{subfigure}{0.49\textwidth}
    \includegraphics[width=\textwidth]{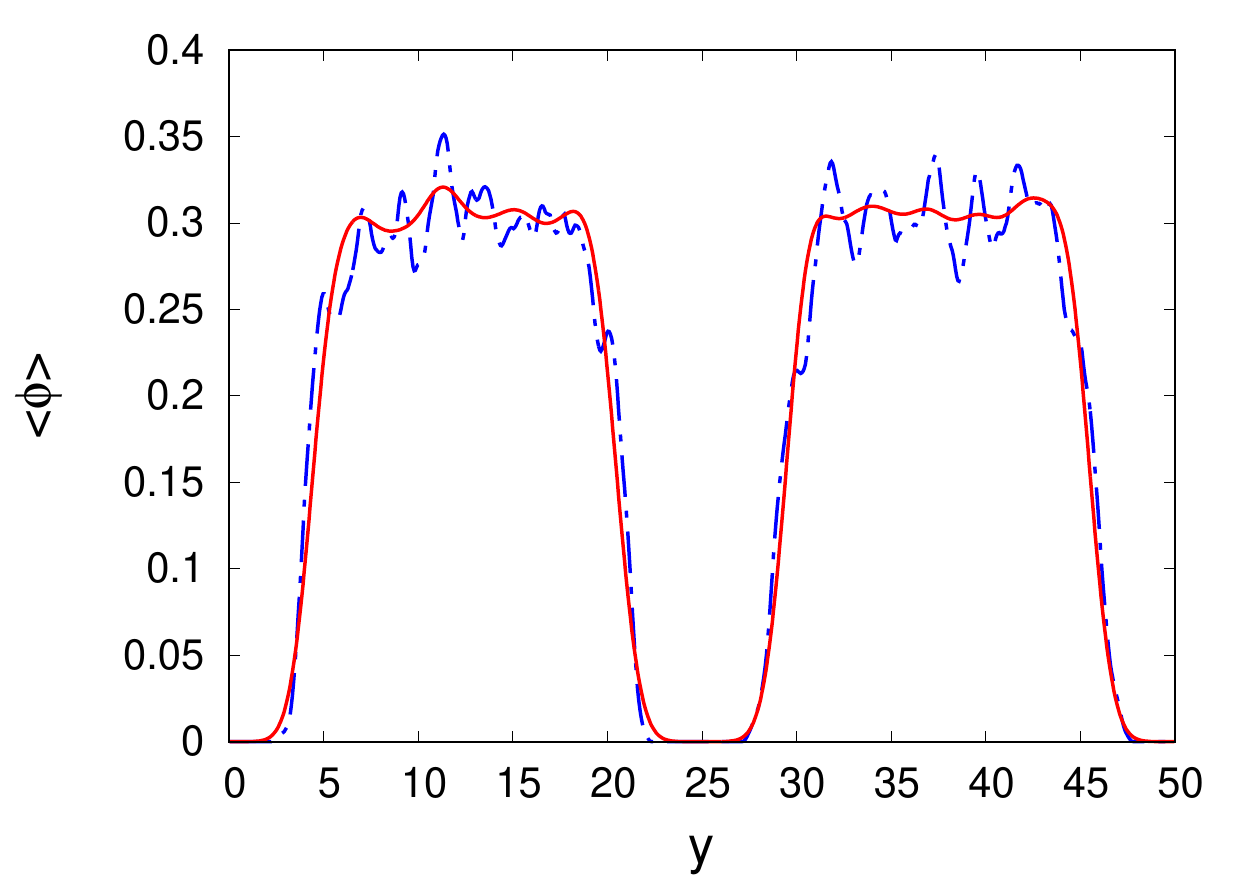}
  \end{subfigure}
  \begin{subfigure}{0.49\textwidth}
   \includegraphics[width=\columnwidth]{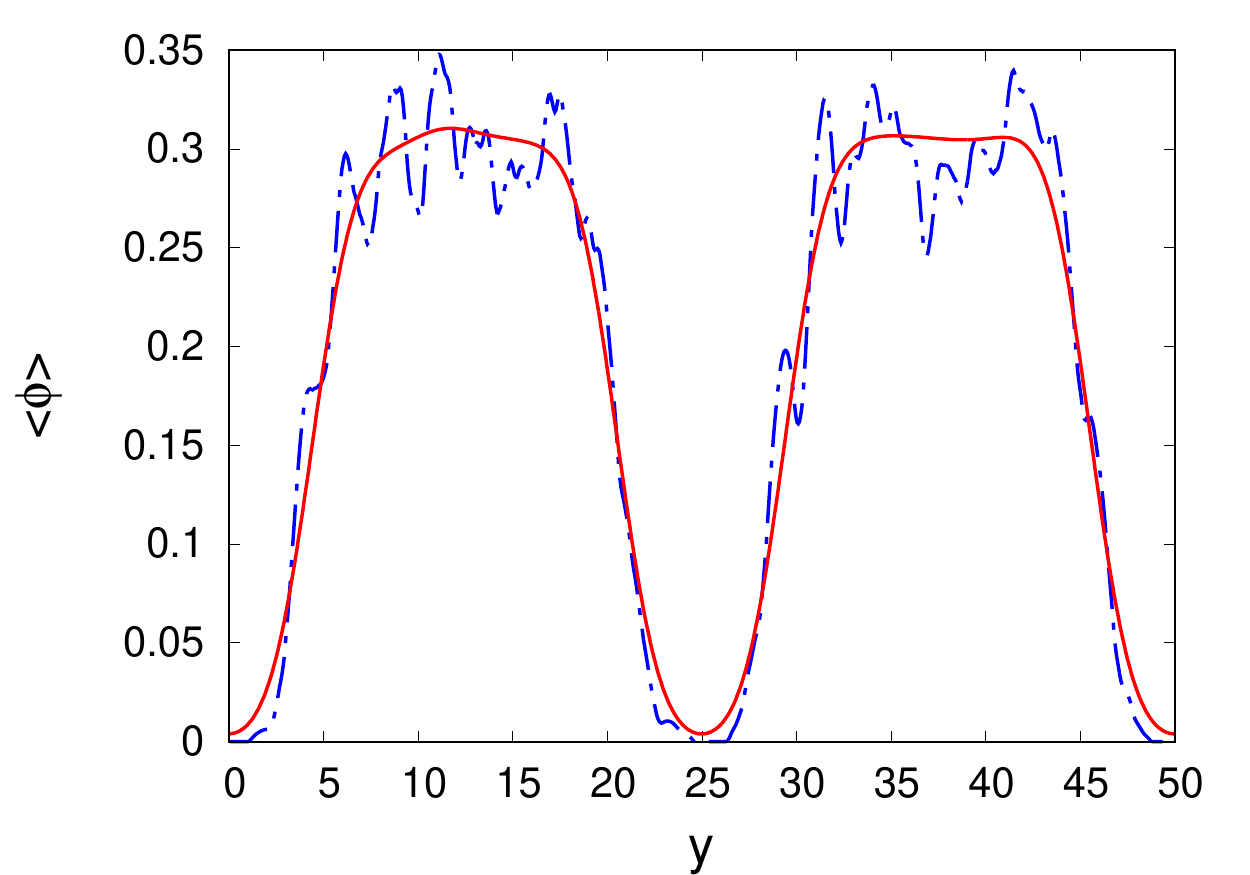}
   \end{subfigure}
  \caption{Comparison between the simulated (dashed-dot blue line) and the approximated (solid red line) average volume fraction $<\phi>$ for the simulation with $\phi = 20$\%. (Top panel) comparison at time $\gamma = 10$; (bottom panel) comparison at time $\gamma = 60$.\label{fig:diffusion_comparison}}
\end{figure}

Finally, we provide an estimate for the droplet diffusion in the vorticity direction, leading to the disappearance of the bands. We consider the diffusion of bands as broadening of the average volume fraction $<\phi>$ (computed by averaging in $x$ and $z$ directions) in the vorticity direction when coalescence is prohibited and droplet pair interaction are more collision-like, as shown in figure \ref{fig:distribution_collision}. As a first approximation, we suppose that the evolution of $<\phi>$ follows a monodimensional diffusion equation with a constant diffusion coefficient
\begin{equation}
  \label{eqn:diffusion}
  \frac{\partial <\phi>}{\partial t} = \mathcal{D}\frac{\partial^2 <\phi>}{\partial y^2}.
\end{equation}
The diffusion coefficient $\mathcal{D}$ has dimension $\mathcal{L}^2/\mathcal{T}$ and can be estimated using as length scale the particle diameter $\mathcal{L} = d$ and as time scale the inverse of the collision frequency per unit volume times the volume associated to the reference length $\mathcal{T} = 1/(d^3C)$. With this definition we find that $\mathcal{D} = (2/3)\dot{\gamma}d^8n^2 = 0.038$. We then solve the diffusion equation \eqref{eqn:diffusion} for several diffusion coefficients starting with an initial distribution of $<\phi>$ corresponding to the initial configuration of our fully resolved simulations (red dashed-dotted line in the inset of figure \ref{fig:bande_merging}). We compute the $\ell^2-norm$ of the error between the approximated distribution given by the diffusion model and the simulated one every 0.01 shear unit and find that the diffusion coefficient which minimize the error is $\mathcal{D} = 0.036$. The difference between the two diffusion coefficients is less then 5\% which can be explained by the deformation of the droplets, not considered in the estimation of the collision frequency $C$ by \citeauthor{SmoluchowskiM.1917}\citep{SmoluchowskiM.1917}. This confirms that the evolution of the average volume fraction $<\phi>$ in the vorticity direction is governed by the timescale of the collision frequency. Finally, we display in figure \ref{fig:diffusion_comparison} the evolution of the approximated average volume fraction $<\phi>$ alongside the one obtained with the full 3D numerical simulations. Since we considered only a purely 1D diffusion equation with a constant diffusion coefficient the evolution of the peaks is not properly reproduced whereas the average diffusion of the band is well approximated.

\section{Conclusions}
We performed numerical simulations of emulsions in a shear flow at moderate volume fractions and low Reynolds number reproducing the experiments in Ref.~\citenum{Caserta2012}. The aim of this study is to demonstrate that coalescence is responsible for a substantial change in the rheology of the suspension and for the formation of the vorticity banding in shear flows of emulsions. Starting with an initial distribution of the disperse phase in banded structures, we observe that the distribution is stable and that the bands remain localized in their initial position, when coalescence is active. In this configuration, the curve of the effective viscosity vs the volume fraction exhibits a negative concavity, as also observed experimentally \citep{Caserta2012}. To single out the effect of the coalescence we introduced a short-range repulsive force which always prevents the merging of drops. When coalescence is prevented, the banded structures are not stable anymore, the droplets tend to diffuse and to assume a uniform distribution across the domain. In this case, the concavity of the curve of the effective viscosity vs volume fraction changes sign, resembling the behavior of suspensions of rigid and deformable particles. The coalescence mechanism allows emulsions to reduce the total surface of the system and hence to reduce the energy dissipation associated to the deformation of the particles. The results of the simulations and band stability indicate therefore that the coalescence is the physical mechanism that allows emulsions to generate the banded structure in the vorticity direction and the collision process is responsible for the timescale of the process.
\begin{figure}
  \centering
  \includegraphics[width=\columnwidth]{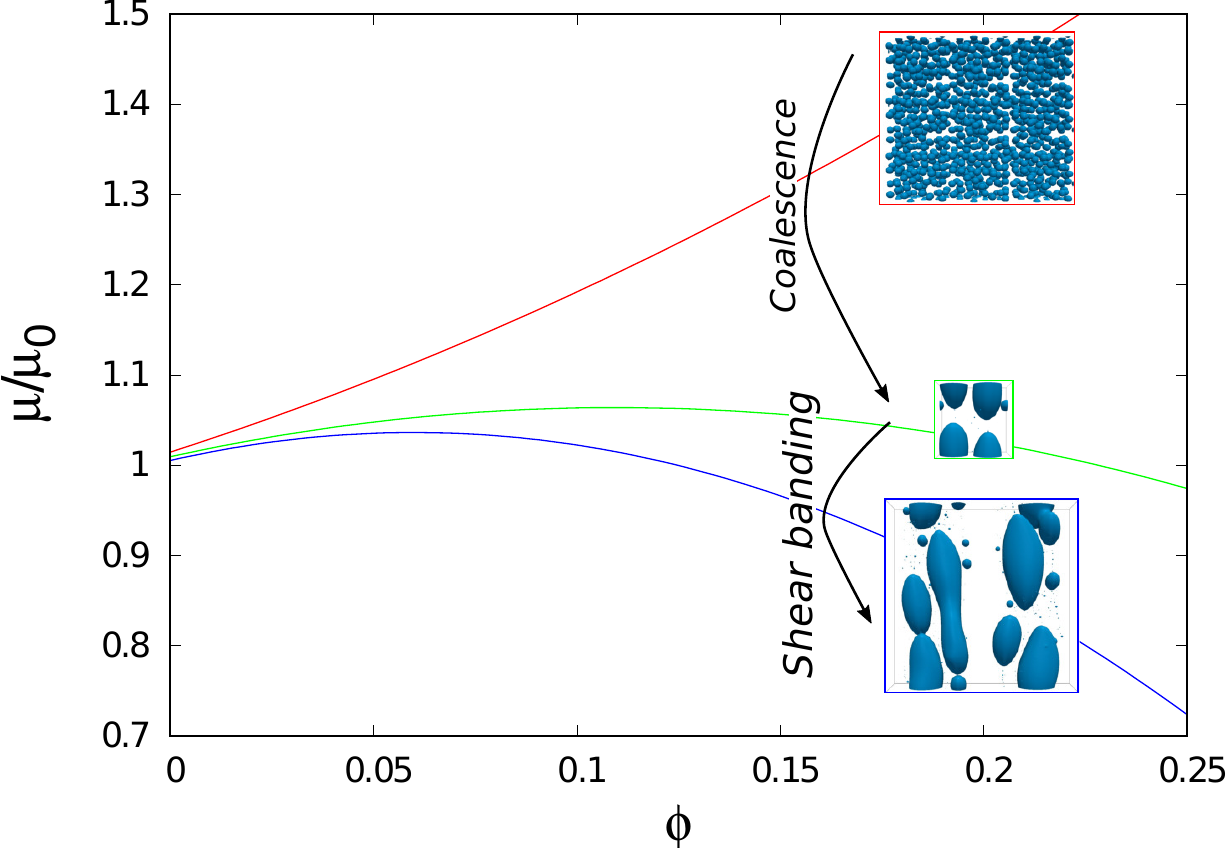}
  \caption{Sketch of the mechanisms contributing to the reduction of the effective viscosity of an emulsion.}
  \label{fig:final_sketch}
\end{figure}
To summarize, there are two mechanisms that allows emulsions to reduce their effective viscosity: \emph{i)} the coalescence, which reduce the total surface of the disperse phase and change the concavity of the rheological curve; \emph{ii)} the shear banding that, when possible in large enough channels, further reduces the interface tension contribution. We sketch qualitatively this behavior of emulsions in figure \ref{fig:final_sketch}.

Future investigations should consider the improvement of the collision algorithm in order to handle collision and coalescence together and reproduce a more realistic collision efficiency.

\section*{Conflicts of interest}
``There are no conflicts to declare''.

\section*{Acknowledgments}
The work is supported by the Microflusa project and the European Research Council grant no. ERC-2013-CoG-616186. The Microflusa project receives funding from the European Union Horizon 2020 research and innovation program under Grant Agreement No. 664823. S. Caserta gratefully thanks Professor Stefano Guido for useful discussions and scientific support. 



\balance


\bibliography{biblio} 
\bibliographystyle{rsc} 

\end{document}